# Can Microsoft Academic assess the early citation impact of in-press articles? A multi-discipline exploratory analysis[1]

Kayvan Kousha, Mike Thelwall and Mahshid Abdoli. Statistical Cybermetrics Research Group, University of Wolverhampton, UK.

Many journals post accepted articles online before they are formally published in an issue. Early citation impact evidence for these articles could be helpful for timely research evaluation and to identify potentially important articles that quickly attract many citations. This article investigates whether Microsoft Academic can help with this task. For over 65,000 Scopus in-press articles from 2016 and 2017 across 26 fields, Microsoft Academic found 2-5 times as many citations as Scopus, depending on year and field. From manual checks of 1,122 Microsoft Academic citations not found in Scopus, Microsoft Academic's citation indexing was faster but not much wider than Scopus for journals. It achieved this by associating citations to preprints with their subsequent in-press versions and by extracting citations from in-press articles. In some fields its coverage of scholarly digital libraries, such as arXiv.org, was also an advantage. Thus, Microsoft Academic seems to be a more comprehensive automatic source of citation counts for in-press articles than Scopus.

## Introduction

Citation indicators derived from conventional scholarly databases, such as Scopus and the Web of Science (WoS), are often used for the impact assessment of published articles. They are rarely useful for recently-published and in-press articles, however, since their citation counts tend to be zero. The overall publication delay (the time between submission or acceptance and publication) also negatively influences citation indicators (Luwel & Moed, 1998; Yu, Wang, & Yu, 2005; Tort, Targino & Amaral, 2012; Shi, Rousseau, Yang, & Li, 2017). Traditional citation indexes seem to wait for articles to be formally published by journals before processing their references. For instance, on 15 October 2017 Scopus had indexed over 277,000 ''In-Press'' articles that had been published as "Online First" or similar in journals. Nevertheless, Scopus does not index or display the cited references of in-press articles until their final version is published in a journal issue (as of 20 October 2017[2]). Hence, it seems likely that millions of citations from in-press articles are not included in any Scopus citation counts. WoS seems to wait for in-press articles to be published in an issue before reporting them. For instance, although on 15 October 2017 Scopus found 52 and 46 in-press articles in 2017 from *Scientometrics* and *Journal of the Association for Information Science and Technology,* respectively, none had been indexed in WoS.

Advance online publication increases citations to articles, Journal Impact Factors and Immediacy Index values (Alves-Silva et al., 2016; Al & Soydal, 2017; Echeverría, Stuart, & Cordón-García, 2017; Todorov & Glänzel, 1988). Many academic publishers provide early online access to their journal articles to minimize publication delays and perhaps to increase citation rates such as *Springer* (Online First), *Wiley* (Early View), *Taylor & Francis* (Latest Articles), and *Nature Publishing Group* (Advance Online Publication). Some authors deposit preprints or postprints (final drafts after peer review) of their articles to open access repositories, such as arXiv.org, or share them via academic social websites, such as ResearchGate or Academia.edu before they are available as online first or final versions in publishers' websites[3]. These strategies reduce article publication delays and presumably make it more likely that an article is cited before it is formally published, especially for journals with long publication backlogs. For example, the article "*Stationary graph processes and spectral estimation*" was published online first in *IEEE Transactions on Signal Processing* on 11 August 2017 and cited 42 times in Google Scholar (16 October 2017) but had not been cited in Scopus or WoS. All 42 citations were to a preprint version of the article that had been deposited in arXiv.org on 14 March 2016. In general, citation rates are influenced by online availability, publication date, and indexing date (Haustein, Bowman, & Costas, 2015).

---

[1] Kousha, K, Thelwall, M. & Abdoli, M. (2018). Can Microsoft Academic assess the early citation impact of in-press articles? A multi-discipline exploratory analysis. Journal of Informetrics, 12(1), 287-298.
[2] https://service.elsevier.com/app/answers/detail/a_id/11241/supporthub/scopus/

From the above discussion, advance online publication is an increasingly important phenomenon that needs to be investigated by scientometricians. This paper assesses the ability of Microsoft Academic to find citations to recently published research by comparing Microsoft Academic citations to over 65,000 in-press articles from 2016 and 2017 with Scopus citations across 26 fields.

## Early citation impact

Early citation impact evidence could help to identify cutting-edge research that quickly attracts citations, differentiating it from typical articles that need longer to be cited (Moed, 2005). Early impact evidence for recent research can also be useful to predict the long-term citation impact of articles. This can support timely research evaluation exercises, academic promotion, the employment of early-career researchers, and the evaluation of research funding programs (Levitt & Thelwall, 2011; Bornmann, 2013; Bruns & Stern, 2016). In justification of these applications, early citation impact (1-2 years after publication) for scientific articles positively correlates with citation indicators calculated in subsequent years (Adams, 2005). Nevertheless, predicting the future citation impact of articles based on early citation counts is challenging due to factors like field differences in citation behaviour (Wang, 2013).

Some alternative sources of evidence have been proposed to identify the early intellectual impact of research, including article downloads (Kurtz et al., 2005; Brody, Harnad, & Carr, 2006; Bollen & Van de Sompel, 2008), Mendeley reader counts (Thelwall & Sud, 2016; Maflahi & Thelwall, 2017) and social web mentions (Thelwall, Haustein, Larivière, & Sugimoto, 2013; Zahedi, Costas, & Wouters, 2014). These all reflect types of use that are likely to appear before citations. Some, such as download counts, may reflect different degrees of interest or uses of academic research compared to citations (Kurtz & Bollen, 2010). Similarly, Mendeley reader counts partly reflect professional, teaching and educational uses (Mohammadi, Thelwall, & Kousha, 2016).

### Free citation indexes for early citation analysis

Several free scholarly websites, including Google Scholar, ResearchGate and Microsoft Academic, index or host preprint versions of articles that could be used for early citation impact assessment.

*Google Scholar*

Google Scholar may be the largest index for the early citation impact of research because it generates higher citation counts than traditional citation databases. It is helpful in this regard by indexing different publishers and wider online sources, such as open access publications (Bar-Ilan, 2008; Khabsa & Giles, 2014; Halevi, Moed, & Bar-Ilan, 2017). For instance, the in-press article "*Stochastic multicriteria decision-making approach based on SMAA-ELECTRE with extended gray numbers*" published online first in the journal *International Transactions in Operational Research* on 7 February 2017 (DOI: 10.1111/itor.12380), had no Scopus citations by 25 September 2017 but had received eight Google Scholar citations from other recently-published journal articles (mostly online first). All eight citing journals found in Google Scholar were covered by Scopus and so indexing delays in Scopus were the reason for its missing citations. On 18 October, Scopus found two of the missing citations to the above article, confirming that indexing delays were the cause. These delays may be for technical (delays in accessing or processing publications) or quality control (waiting for the version of record) reasons. Despite the substantial Google Scholar coverage of scholarly publications and citations, it cannot be used for most research evaluations because it does not allow automatic data collection, which is a practical necessity for large scale analyses. The Publish or Perish software can extract Google Scholar citations and other citation impact indicators for individual papers, academics or journals, however (Harzing, 2007).

---
[3] http://www.sherpa.ac.uk/romeo/statistics.php?colour=green

*ResearchGate*

ResearchGate reports citation counts on its article profile pages by extracting citations from other publications uploaded to the site. It is therefore both a citation index and a digital repository. ResearchGate finds more citations to recently-published library and information science articles than both WoS and Scopus (Thelwall & Kousha, 2017). Nevertheless, it does not allow automatic data collection and so it is not a practical data source for large scale citation analyses.

*Microsoft Academic*

The new scholarly publication index Microsoft Academic was officially launched in July 2017, replacing an unsuccessful predecessor. Microsoft Academic claims to include records for over 170 million scholarly publications from publisher websites, authors' personal homepages and documents indexed by the Bing search engine. It also allows free API searching for a limited number of queries per month (https://academic.microsoft.com/). Recent studies suggest that Microsoft Academic could be a useful substitute for a large-scale citation impact assessment in the absence of any Google Scholar automatic search capability (Harzing & Alakangas, 2017a; Hug, Ochsner, & Brändle, 2017).

A comparison between citation counts from Microsoft Academic, Google Scholar, Scopus and WoS to a sample of publications from 145 University of Melbourne academics in five broad fields showed that Microsoft Academic found more citations than Scopus and WoS in Engineering, Social Sciences, and the Humanities, but citation counts for the Life Sciences and the Sciences were similar across all four databases. Although Google Scholar returned the highest citation counts across all fields, the average monthly growth of Microsoft Academic citations for a 5.5-month period was higher than the other databases (Harzing & Alakangas, 2017a). Seven months after the initial study, Microsoft Academic citation counts for the same Australian publications were similar to those from Google Scholar for science fields, but much lower for the Social Sciences and Humanities due to lower coverage of non-journal publications in Microsoft Academic compared with Google Scholar (Harzing & Alakangas, 2017b). The above studies suggest that Microsoft Academic has wide but variable coverage of academic research.

Two investigations have compared Scopus and Microsoft Academic citations from selected journals in different subject areas. A study of articles published in *Nature*, *Science* and seven library and information science journals from 1996 to mid-2017 found similar average citation counts between Microsoft

Academic and Scopus, with no obvious early citation advantage for Microsoft Academic (Thelwall, 2017a). Nevertheless, a larger scale investigation with over 172,000 articles published 2007-2017 in 29 journals found that overall Microsoft Academic captured slightly more (6%) citations than Scopus. The difference was much higher (51%) for the recently-published articles from 2017 (Thelwall, 2017b).

Microsoft Academic indexes some types of document that are not included in Scopus and WoS, such as working papers and theses (Hug & Brändle, 2017). This gives it the potential to report higher citation counts than WoS and Scopus and suggests that in some fields the nature of the impact reflected in its citation counts may be different. Microsoft Academic also indexes papers from some preprint archives, such as the Social Science Research Network (SSRN) and arXiv (Thelwall, submitted). It is not clear how comprehensive Microsoft Academic's coverage of individual preprint archives is, nor whether it indexes all open access online preprint archives but this indexing gives it the potential to identify early citations from and to preprints before publication. Microsoft Academic does not index all open access preprint archives comprehensively (Hug & Brändle, 2017), perhaps because it cannot parse all PDF formats.

All the above results were based on publications derived from selected academics or journals and did not focus on in-press articles. Hence, more systematic evidence based on a random sample of advance online publications is needed to assess Microsoft Academic for monitoring the early citation impact of in-press research.

## Research questions

This paper addresses the above-mentioned gap in knowledge about the value of Microsoft Academic for in-press articles. As discussed above, conventional citation databases do not seem to reflect the early citation impact of research and Google Scholar searches cannot be automated for large-scale analyses. Thus, Microsoft Academic, which supports API searching and automatic data collection, may be suitable for this. Despite several recent studies of Microsoft Academic, it is not known whether it captures substantial numbers of early citations to in-press articles, and if so, what types of citing sources it identifies. The following research questions drive this study.

1. Does Microsoft Academic identify more citations than Scopus to in-press articles?
2. Do Microsoft Academic citations to in-press articles reflect a similar type of impact to Scopus citations?
3. Which types of document does Microsoft Academic find that cite in-press articles?

## Methods

Scopus and Microsoft Academic were used to identify citations to in-press articles across all 26 broad Scopus fields. In-press articles from 2016 and 2017 were chosen as the most recent publication type registered in Scopus. Both years were included so that the influence of the first publication year could be assessed. Scopus claims that "Articles in-press are documents that have been accepted for publication, but have not yet been assigned to a journal issue" and "the pre-published versions of documents available online within four days after being transferred from publishers"[4]. Both Scopus and Microsoft Academic citations were gathered for each field and year during 23-29 September 2017. The publication year of an in-press article in Scopus refers to the date when it was published in early view or online first rather than the date of publication in a journal issue.

---
[4] https://service.elsevier.com/app/answers/detail/a_id/11241/supporthub/scopus/

## Scopus data set

The search command *DOCTYPE(ip)* was used in the advanced Scopus search interface to retrieve all articles with *In-Press* status. The results were restricted to each of the 26 broad fields and years (2016-2017) separately and were also restricted to journals (excluding Book Series, Trade Publications and Conference Proceedings). The Scopus email export feature was used to extract the maximum number of articles (20,000) for each field and year. For searches with over 20,000 results (Medicine, Engineering and Social Science) the results were limited to different countries to return less than the maximum of 20,000 records for download in each chunk. Scopus records without authors ("[No author name available]"), titles ("[No title available]") or DOIs were excluded and articles with less than three words in their titles were also excluded from the data set because in most cases they were non-article documents (e.g., Preface, Foreword, Editor's Comments, In this issue or In memoriam) and could generate false matches via Microsoft Academic title searches (see below). For each field, random samples were taken from articles in-press published in 2016 (1,000 articles) and 2017 (1,500 articles). These sample sizes were chosen to restrict the cost of the Microsoft Academic automatic queries whilst giving large enough samples to examine differences between fields. More random samples were taken from 2017 for each field because there were more in-press articles in 2017 (about 277,000) than 2016 (about 98,000) in Scopus on 15 October 2017. More articles were also included in the random samples for broad fields with many in-press articles in Scopus: Medicine, Engineering, Biochemistry, Genetics and Molecular Biology, and Social Science (Table 1) because about half of the in-press articles in Scopus were from these four fields (Medicine about 20%, Engineering 11%, Social Science 9% and Biochemistry, Genetics and Molecular Biology 8%).

## Automatic Microsoft Academic citation searches

The free software *Webometric Analyst* (http://lexiurl.wlv.ac.uk) was used to generate and perform automatic searches via the Microsoft Academic API. The Microsoft Academic queries were automatically generated from the Scopus data (see "Microsoft Academic – Make queries" options in "Citations" menu) based on searching article titles. The software uses different algorithms to converts characters in article titles in accordance with Microsoft Academic's indexing strategy (see Thelwall, 2017b). A title search strategy was used because the recall and precision of Microsoft Academic API searches by articles titles with subsequent DOI-based filtering are very high for individual journal articles (Thelwall, 2018). Below are examples of queries used for Microsoft Academic API searches by articles title.

*Ti='bibliometric profile of deep brain stimulation'*
*Ti='medical mistrust and colorectal cancer screening among african americans'*

The queries were run using Webometric Analyst ("Microsoft Academic – Run queries") via a free trial key for the Microsoft Academic API. The results were matched with Scopus based on their DOIs to give precise matches, using a feature designed for this purpose (Microsoft Academic – "Filter results"). This extra step is necessary because Microsoft Academic returns some false matches and currently lacks a DOI search capability. Scopus articles that were not found in Microsoft Academic using their DOIs were excluded from the study, giving 40,808 matching records out of 45,000 for 2017 (91%) and 24,779 out of 26,987 for 2016 (92%) (Table 1). Taking a conservative approach and assigning zero citations to the unmatched articles in

Microsoft Academic and keeping their Scopus citations (which would be mostly 0) would not change the results substantially, as can be seen below.

## Data analysis

Geometric means were used to compare the average number of Scopus and Microsoft Academic citations per article because the arithmetic mean is a less precise central tendency metric for highly skewed data. The median is also appropriate for skewed data but was not effective to discriminate between citation counts from the two sources due to many zeros and ones in the data sets. The percentage of articles with at least one citation was also used to compare Scopus and Microsoft Academic from a different perspective (Thelwall, 2016). Confidence intervals were calculated using the geometric mean formula (Fairclough & Thelwall, 2015).

Correlations are a suitable technique to compare data sources to assess whether they reflect a similar type of impact, although they provide only partial evidence (Sud & Thelwall, 2014). Due to the skewed distribution of citation counts, Spearman correlations were used instead of Pearson correlations to assess the association between Scopus and Microsoft Academic citations. Correlations were calculated independently for each of the 26 fields and years (2016-2017) to make the results more meaningful since merging fields or years can artificially inflate correlation coefficients (Fairclough & Thelwall, 2015).

The unique sources of citations found by Microsoft Academic were manually checked to identify whether they were present in Scopus. The purpose was to investigate whether Microsoft Academic had indexed non-Scopus journals, books, or research reports or whether the higher number of Microsoft Academic citations compared with Scopus was due to capturing early citations from publications covered by Scopus. Ten articles with the most Microsoft Academic citations but no Scopus citations were selected from the set of articles published in 2017 for each of the 26 fields (n=260 articles). The 260 selected articles were cited by 1,122 publications, as reported by Microsoft Academic during 23-29 September 2017. To answer the third research question, three characteristics of the 1,122 citing documents were manually checked as shown below. Publisher websites were checked to identify the language and citing source type when they were not recognizable from the Microsoft Academic search results. Extra searches were sometimes also conducted using Google or Google Scholar to identify characteristics of Microsoft Academic citations.

- **Citing source type:** *Journal article*; *Conference paper*; *Book/book chapter*; *Thesis*; *E-print* or *Other*. This category may reveal if one type of publication origin is less common in Scopus compared to Microsoft Academic.
- **Language of citing source:** *English* or *non-English*. Language bias is an obvious facet to investigate, given the dominance of English in Scopus.
- **Citing source publication year:** Publication years were recorded to investigate whether citations to in-press articles in 2017 were from other publications in the same year (2017) or whether Microsoft Academic had captured citations from publications before 2017 to preprint versions of in-press articles.

# Results

## RQ1: Microsoft Academic citations vs. Scopus citations

### Geometric means of Scopus and Microsoft Academic citations

Microsoft Academic found more citations than did Scopus for in-press articles across all 26 Scopus broad fields from 2017 and 2016 (Figures 1 and 2). Because the confidence intervals do not overlap (except for Dentistry in 2017), the difference between Microsoft Academic and Scopus average citation counts is statistically significant at the 95% level.

In Mathematics, the geometric mean number of Microsoft Academic citations to 2017 in-press articles (0.158) is over ten times more than for Scopus citations (0.016). In five social science and the arts and humanities Scopus fields the geometric mean number of Microsoft Academic citations is up to five times higher than Scopus citations: Decision Science (5.4); Arts and Humanities (4.6); Business (4.5); Social Science (4.0); Economics (3.6). This difference is lower for most science and medical science fields (about 2-3 times), except for engineering (3.6).

These results suggest that Microsoft Academic citations could be an especially useful early impact indicator in the social sciences, arts and humanities. This supports a previous finding that Microsoft Academic identified 1.5 to 2 times more citations than Scopus and WoS to publications by Australian academics in the social sciences (Harzing & Alakangas, 2017a). Nevertheless, the ratio of the geometric mean number for Microsoft Academic to Scopus citation for 2016 in-press articles is much lower (e.g., about 3 times higher than Scopus citation counts in Mathematics) than for the 2017 data set. This suggests that Scopus can add more citations to recent articles one year after their online first publication, although not as many as Microsoft Academic (Figure 2).

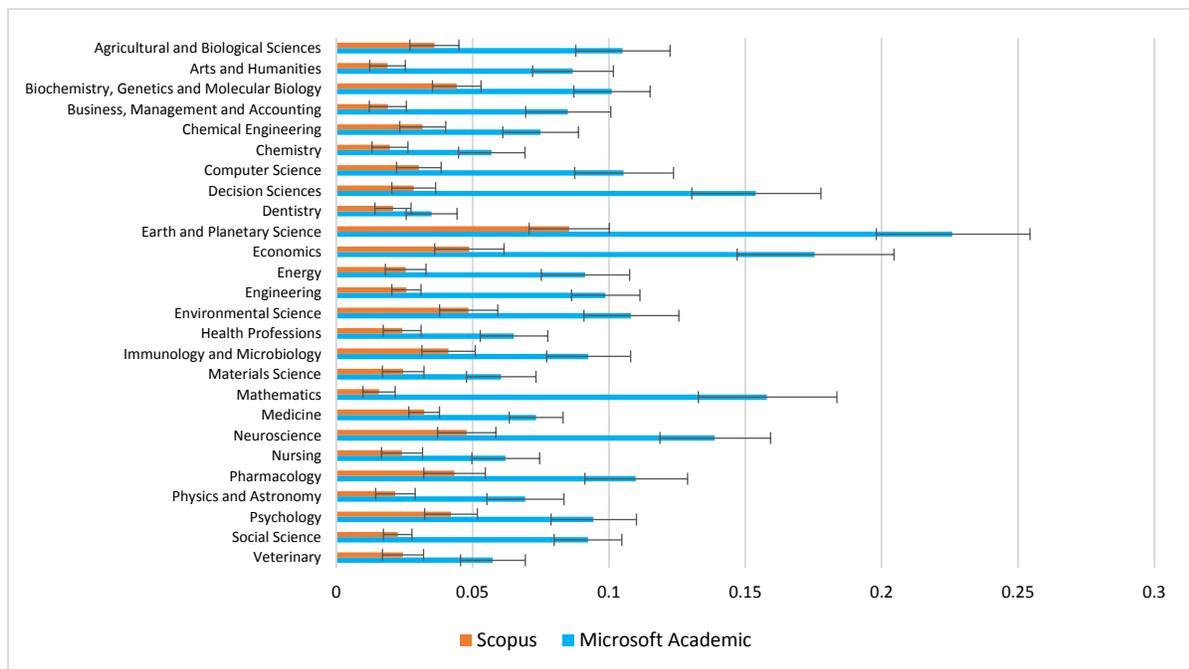

Figure 1. Geometric mean number of Scopus and Microsoft Academic citations and 95% confidence intervals for 2017 in-press articles across 26 fields.

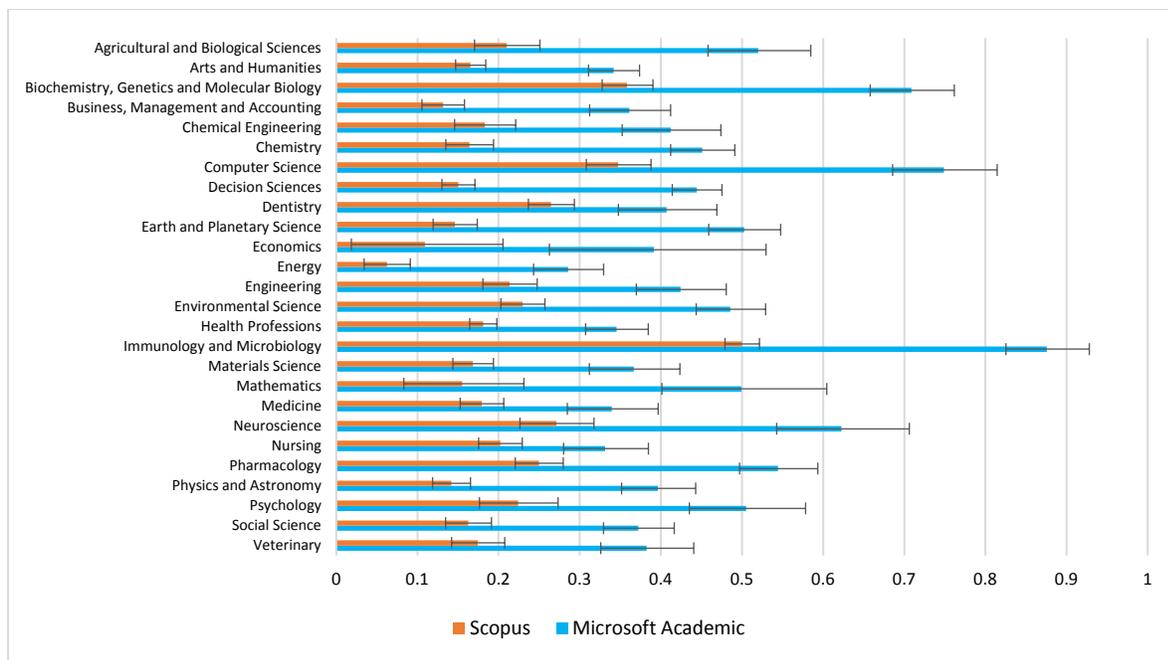

Figure 2. Geometric mean number of Scopus and Microsoft Academic citations and 95% confidence intervals for 2016 in-press articles across 26 fields.

### Proportion (non-zero) of cited articles in Scopus and Microsoft Academic

Because most in-press articles do not have enough time to be cited by academic publications, the proportion of cited papers is an alternative method for assessing the statistical significance of differences between Scopus and Microsoft Academic citations to recent articles in a specific field and year (Thelwall, 2017c).

In general, higher proportions of recently-published articles were cited by Microsoft Academic than Scopus and this difference is statistically significant at the 95% level across all Scopus broad fields, except for Dentistry in 2017 (Figures 3 and 4). Obviously, lower proportions of articles in-press from 2017 have been cited than articles from 2016 because they had less time to be cited by other publications. In Mathematics, the proportion of articles from 2017 that had one or more citations in Microsoft Academic is 7 times higher than for Scopus. In the Arts and Humanities, Business, Decision Sciences and Social Science, Microsoft Academic found 4 times more articles from 2017 than Scopus with at least one citation. In other science, engineering and medical fields, Microsoft Academic found 1.5-3 times more articles with at least one citation than Scopus (e.g., Engineering: 3.1 and Dentistry 1.5 times higher than Scopus). Although the proportion of in-press articles from 2016 with one or more Microsoft Academic citations is still higher than for Scopus citations across all fields (1.3-2.5 times higher than Scopus), this difference is lower compared with 2017 articles, probably because Scopus had more time to add citations for older articles.

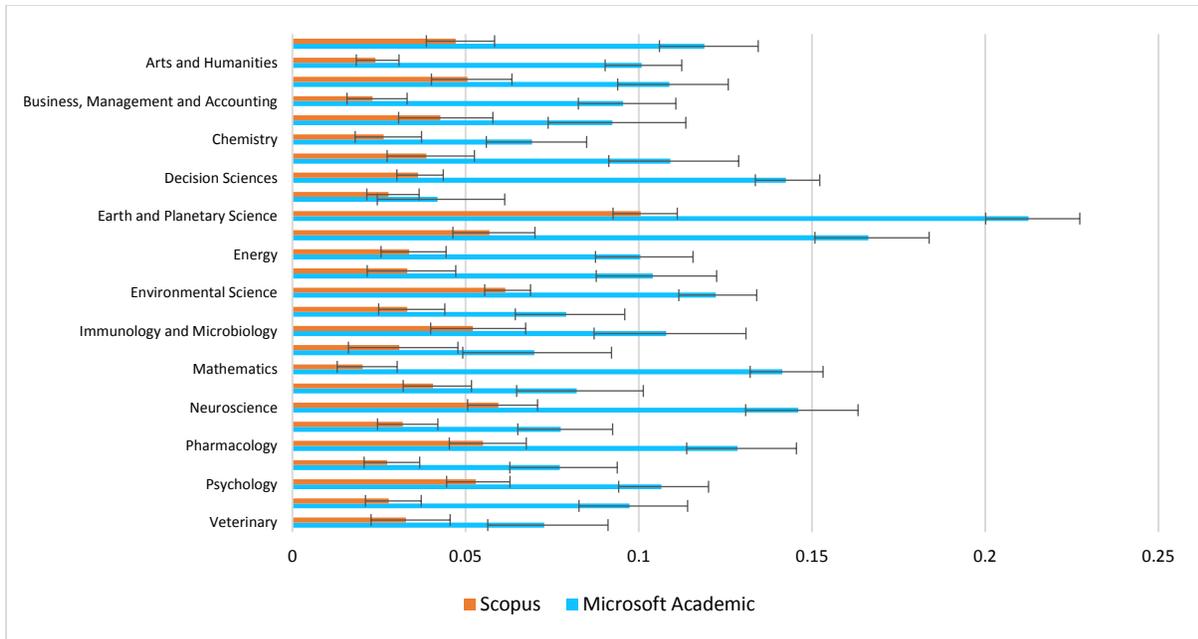

Figure 3. Proportion of articles with at least one citation in Scopus and Microsoft Academic and 95% confidence intervals for 2017 in-press articles across 26 fields.

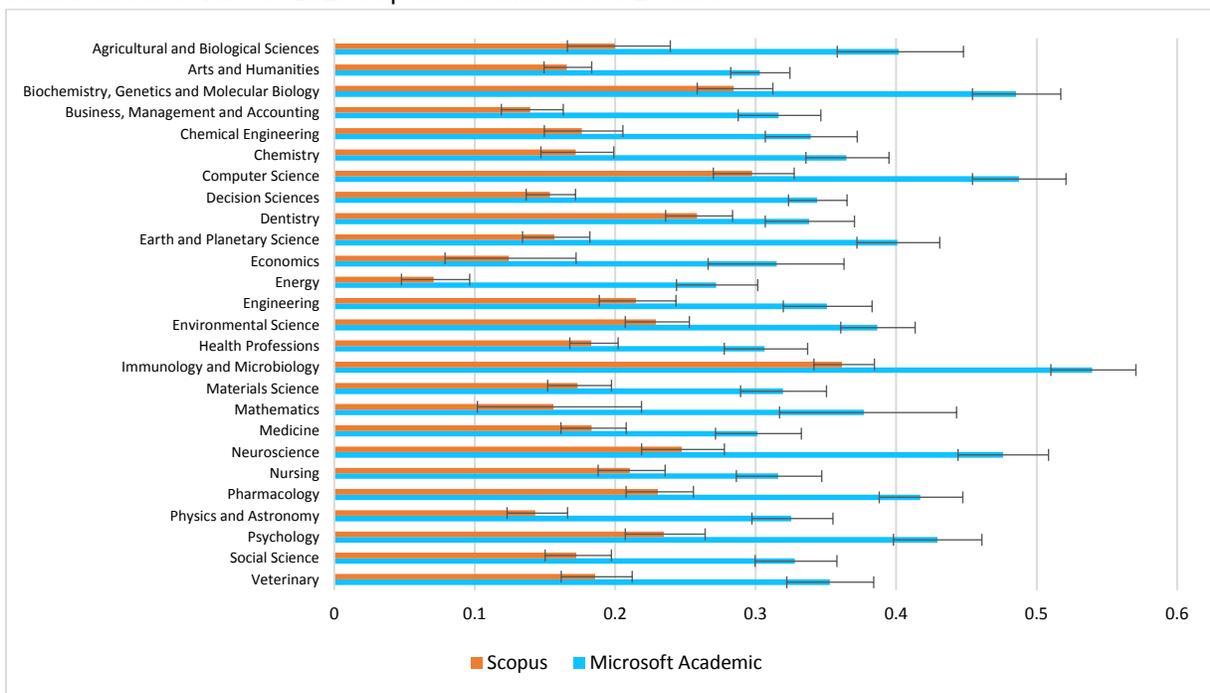

Figure 4. Proportion of articles with at least one citation in Scopus and Microsoft Academic and 95% confidence intervals for 2016 in-press articles across 26 fields.

### RQ2: Correlations between Scopus and Microsoft Academic citations

There are significant moderate positive Spearman correlations between the Scopus and Microsoft Academic citation counts in most fields for articles in-press from 2017 (Table 1). The correlations for

Mathematics (.298), the Arts and Humanities (.383), Decision Sciences (.384), Business (.396) and Social Sciences (.398) are lower than for other fields, perhaps because Scopus found few citations in these fields, as shown in Figure 1 and Figure 3 (see also: Thelwall, 2016). Correlations are higher in all science fields (about 0.5-0.6), indicating that both databases reflect similar patterns of citation impact for in-press articles. The correlations between the Scopus and Microsoft Academic citation counts are higher for the 2016 data set of articles in most fields (expect for Veterinary, Earth and Planetary Science, and Energy), reflecting the higher citation counts from both sources (higher averages tend to increase correlations: Thelwall, 2016). For instance, the correlations increased from 0.298 to 0.507 in Mathematics, and from 0.383 to 0.600 in the Arts and Humanities and from 0.403 to 0.582 in Social Sciences from 2017 to 2016.

Table 1. Spearman correlations between Scopus and Microsoft Academic citation counts for in-press articles in 2017 and 2016 across 26 fields.

| Scopus fields | 2017 in-press articles | | 2016 in-press articles | |
| --- | --- | --- | --- | --- |
| | Articles matched with DOIs (articles in the sample) | Spearman correlation | Articles matched with DOIs (articles in the sample) | Spearman correlation |
| Agricultural and Bio Sci. | 1337 (1500) | .574** | 961 (1000) | .591** |
| Arts and Humanities | 1419 (1500) | .383** | 954 (1000) | .600** |
| Biochemistry | 2215 (2500) | .588** | 964 (1000) | .664** |
| Business | 1383 (1500) | .396** | 996 (1000) | .562** |
| Chemical Eng. | 1288 (1500) | .585** | 970 (1000) | .655** |
| Chemistry | 1329 (1500) | .515** | 960 (1000) | .590** |
| Computer Sci. | 1421 (1500) | .505** | 921 (1000) | .650** |
| Decision Sci. | 1404 (1500) | .384** | 925 (1000) | .535** |
| Dentistry | 1408 (1500) | .505** | 213 (403[+]) | .680** |
| Earth and Planetary Sci. | 1393 (1500) | .574** | 983 (1000) | .503** |
| Economics | 1107 (1500) | .488** | 902 (1000) | .503** |
| Energy | 1394 (1500) | .493** | 861 (1000) | .404** |
| Engineering | 2836 (3000) | .501** | 1243 (1500) | .637** |
| Environmental Sci. | 1366 (1500) | .638** | 908 (1000) | .640** |
| Health Professions | 1417 (1500) | .502** | 973 (1000) | .596** |
| Immun. and Microbio. | 1363 (1500) | .578** | 404 (619[+]) | .721** |
| Materials Sci. | 1360 (1500) | .557** | 958 (1000) | .628** |
| Mathematics | 1379 (1500) | .295** | 891 (1000) | .507** |
| Medical Sci. | 3352 (3500) | .495** | 1862 (2000) | .614** |
| Neuroscience | 1363 (1500) | .573** | 861 (1000) | .616** |
| Nursing | 1316 (1500) | .519** | 946 (1000) | .630** |
| Pharmacology | 1090 (1500) | .604** | 868 (1000) | .620** |
| Physics and Astron. | 1387 (1500) | .541** | 971 (1000) | .538** |
| Psychology | 1437 (1500) | .522** | 950 (1000) | .579** |
| Social Sci. | 2764 (3000) | .398** | 1903 (2000) | .582** |
| Veterinary | 1280 (1500) | .612** | 431 (465[+]) | .489** |

** Significant at the p = 0.01 level. +The total number of articles in Scopus was less than the default sample size.

### RQ3: Sources of citations Microsoft Academic but not Scopus

Manual checks were conducted for 1,122 citations to 260 in-press articles published in 2017 with the most Microsoft Academic citations but no Scopus citations across 26 fields. Over three quarters (78%) of the citations found by Microsoft Academic were journal articles, 13% were preprints, 7% were conference papers and 2% were other types of publication (e.g., books and thesis). About three quarters (73%) were published in 2017 and just over a quarter (27%) before 2017. Almost all (99%) were in English.

### Citing source type

Journal articles were the most common type of citing document indexed by Microsoft Academic but not Scopus in most fields and only two fields (Engineering and Mathematics) had less than 50% of their citations from journal articles (Figure 5). There were 504 unique citing journals identified by the manual checks. Matching these against the official Scopus Source List of indexed journals[5], most (91% or 459) were indexed by Scopus (e.g., "*Physics of Life Reviews*" or "*Journal of Ethnic and Migration Studies*") and only 9% (44 journals) were not covered by Scopus at the time of this study (e.g., "*Current Tropical Medicine Reports*" or "*International Journal of STEM Education*"). Thus, Microsoft Academic had captured many citations from Scopus indexed journals before Scopus. Hence, Microsoft's main advantage over Scopus for in-press articles is its citation indexing speed rather than wider coverage of journals.

Microsoft Academic captured many citations from conference papers, especially in Engineering (34%), Computer Science (22%) and Mathematics (19%), because conference papers are important scientific outputs in these fields. For example, there is evidence that over 40% of citations to highly cited publications in computer science are from proceedings papers (Bar-Ilan, 2010). Manual checks of the citations from conference papers found that Scopus had indexed 82% of the conference proceedings of the conference paper citations found by Microsoft Academic (e.g., "*International Conference on Machine Learning and Cybernetics*") and only 18% were outside Scopus (e.g., "*International Congress on Mathematical Software*"). Hence, a combination of Scopus indexing delays for recently published conference proceedings and the extra Microsoft Academic coverage of online conference papers was the main reasons for the missed Scopus conference citations to in-press articles.

Substantial numbers of Microsoft Academic citations were from e-print archives in some subject areas. In Decision Science (41%), Physics and Astronomy (36%), Mathematics (32%) and Computer Science (20%) many Microsoft Academic unique sources of citations were preprints or post prints of articles deposited in arXiv.org. For example, the in-press article "*Density Level Sets: Asymptotics, Inference, and Visualization*" (doi:10.1080/01621459.2016.1228536) to be published in *Journal of the American Statistical Association* had no Scopus citations but had received six Microsoft Academic citations, all from arXiv.org. This suggests that Microsoft Academic returns many citations from key e-print repositories, increasing its early citation counts for recent publications in fields with strong preprint cultures.

---

[5] https://www.scopus.com/source/browse.url

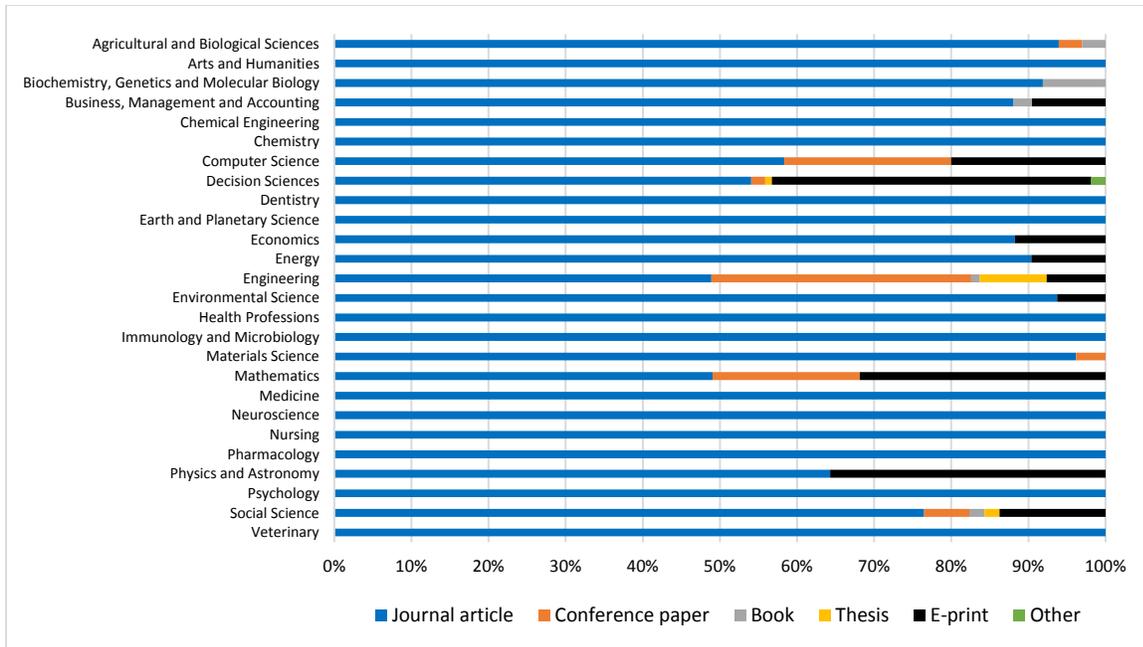

Figure 5. Origins of citations found by Microsoft Academic but not Scopus to 260 in-press articles in 2017 across 26 fields (n=1,122).

### *Citing source publication year*

Most (73%) of the Microsoft Academic citations to in-press articles were published in 2017. Over a quarter (27%) were from citing documents published at least one year before formal publication/acceptance of the in-press articles, as recorded in Scopus: including 15% from 2016, 7% from 2015, and 5% prior to 2015. There are clear disciplinary differences, however (Figure 6). In Decision Science (73%), Mathematics (61%) and Social Science (51%) most MA citations to in-press articles in 2017 were from publications before 2017, suggesting that preprint versions of in-press articles could attract early citations found by Microsoft Academic. This is possible because some authors had previously shared preprints of their in-press articles via e-print repositories. Manual checks showed that Microsoft Academic aggregates citations extensively from *arXiv.org* as well as biorxiv.org (biology and bioscience) and *ssrn.com* (social sciences and humanities).

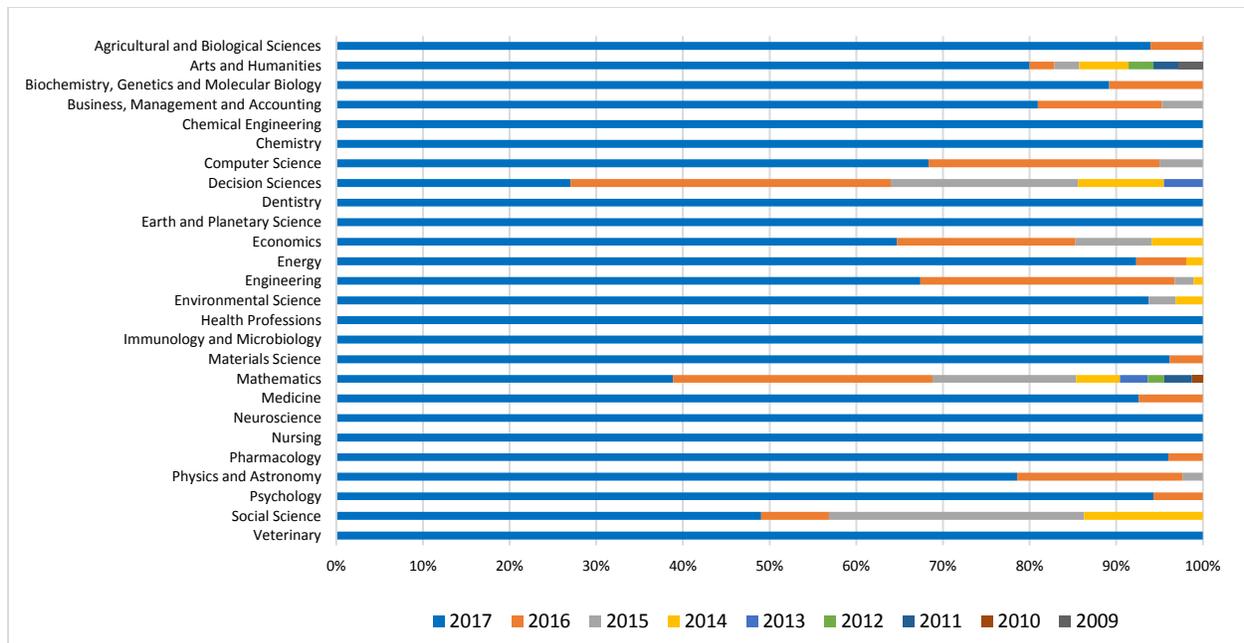

Figure 6. Publication year of citations found by Microsoft Academic but not Scopus for 260 in-press articles in 2017 across 26 fields (n=1,122 citations).

## Discussion

Although Microsoft Academic is known to locate slightly more citations than conventional citation databases overall, depending on discipline, and considerably more for recently-published articles (Harzing & Alakangas, 2017a; Harzing & Alakangas, 2017b; Thelwall, 2017a; Thelwall, 2017b), the results from this study show for the first time that Microsoft Academic citation counts are much higher than Scopus citation counts for in-press articles. Manual checks of 1,122 Microsoft Academic citations not found in Scopus also extend previous results by identifying the main reasons for this difference.

- **Citation indexing speed:** In most fields, the citation advantage of Microsoft Academic is due to identifying early citations to in-press articles from sources indexed by Scopus (91% of journals) which Scopus apparently needs more time to extract citations from. In contrast to Microsoft Academic, which can access cited references from in-press articles as citations to other in-press articles, Scopus does not use or report citations to in-press articles from other in-press articles until article final versions are published in a journal issue. Hence, the faster citation indexing of Microsoft Academic partly explains finding more citations from Microsoft Academic searches than Scopus to the in-press articles.
- **Citations from preprints:** Although Microsoft Academic did not seem to have substantially broader coverage of academic journals than Scopus, it includes many citations from the arXiv.org e-print archive. The manual checks showed that in Decision Science, Physics and Astronomy, Mathematics and Computer Science, up to 40% of the Microsoft Academic citations not found in Scopus were preprints or post prints of articles mostly deposited in arXiv. Scopus does not index arXiv, perhaps for quality control reasons, but this may undermine its effectiveness for early citation counts in preprint-oriented fields like physics and astronomy, mathematics and computer science.

- **Matching preprints with in-press versions of articles:** Microsoft Academic seems to include citations to journal articles that target preprint versions of the same paper. This seems to be an important feature for early citation analysis because there is evidence that early view for articles in ArXiv accelerates the citation rate for the subsequent journal articles (Moed, 2007). For instance, the in-press article "*Partial facial reduction: Simplified, equivalent SDPs via approximations of the PSD cone*" to be published in *Mathematical Programming* (doi:10.1007/s10107-017-1169-9) had no Scopus citations but had 18 Microsoft Academic citations, all to the preprint version deposited in arXiv.org in 2014 (e.g., *F. Permenter and P. Parrilo, Partial facial reduction: simplified, equivalent SDPs via approximations of the PSD cone, arXiv preprint arXiv:1408.4685, 2014*). Microsoft Academic presumably matches author and title information in the cited references of online documents with bibliographic information from publishers (Sinha, Shen, Song, Ma, Eide, Hsu, & Wang, 2015). Hence, if a preprint of an article is cited before the version of record appears in a journal, Microsoft Academic could assign this citation to the journal version.

**Limitations**

For practical reasons, a few of the Scopus in-press articles from 2017 (9%) and 2016 (8%) were excluded when matching Scopus with Microsoft Academic. Only results from Microsoft Academic with DOIs were matched with Scopus to ensure that the matches were correct. Moreover, to answer RQ3, only 260 articles with the most Microsoft Academic citations but no Scopus citations were selected (10 for each 26 fields in 2017), giving 1,122 Microsoft Academic citations outside Scopus. Although Microsoft Academic did not seem to have substantially broader or more international coverage than Scopus, it is possible that there are pockets of higher coverage in some areas which were not found in this sample.

## Conclusions

Since Microsoft Academic identifies substantially more citations to in-press articles than Scopus, it may be useful to assist with evaluations of recent research outputs. The citation advantage of Microsoft Academic has disciplinary variations but is present in some form for all fields analysed. Both the average number (geometric mean) of citations and proportion of cited articles were higher in Microsoft Academic than Scopus in all 26 fields and both years (2016-2017). The differences between Microsoft Academic and Scopus citation counts were statistically significant, except for Dentistry in 2017.

In answer to the second research question, there are significant ($p<0.01$) moderate Spearman correlations between the Scopus and Microsoft Academic citation counts for in-press articles in 2017 across most science, engineering and medical science subject areas, suggesting that both citation databases reflect similar kinds of scientific impact. However, the correlations in Mathematics, the Arts and Humanities, Decision Science, Business and Social Sciences are lower than in other fields (less than 0.4), perhaps because Scopus captures very few citations in these subject areas. Low numbers can reduce the strength of a correlation coefficient for discrete data (Thelwall, 2016). For most of the 26 fields the correlations between Scopus and Microsoft Academic citations for 2016 articles are stronger than those for the 2017 articles, presumably because Scopus had more time to find citations. Thus, the results are consistent with Microsoft Academic's citation counts reflecting a similar type of impact to Scopus which might be helpful to predict the future citation impact of articles.

In answer to the third research question, Microsoft Academic finds more citations than Scopus to in-press articles because its citation indexing speed is faster, it covers e-print archives and it includes citations to journal articles that cite preprint versions (e.g., in arXiv). Only 1% of the extra citations were from non-English scholarly publications, suggesting that it may have poor coverage of non-English publications. This should not be exaggerated, however, since the original data set was extracted from Scopus, which has its own international biases.

Finally, because WoS and Scopus do not index the cited references of in-press articles until their final version is published in a journal issue and Google Scholar does not allow systematic data collection, Microsoft Academic seems to be the most useful automatic tool for assessing the early citation impact of in-press articles for large-scale analyses, especially for recently published, accepted or in-press articles (e.g., 1-2 years after publication). Nevertheless, Microsoft Academic's indexing of un-reviewed archives is also a quality control issue for research evaluations because the citation counts can be gamed by uploading low quality preprints to digital libraries (Sorokina, Gehrke, Warner, & Ginsparg, 2006). In consequence, it should not be used for formal research evaluations when stakeholders are informed in advance (e.g., Wouters & Costas, 2012) but it can be used for other purposes, such as self-evaluations by researchers and research funders (Dinsmore, Allen, & Dolby, 2014).